\documentclass[conference]{IEEEtran}
\ifCLASSINFOpdf
  \usepackage[pdftex]{graphicx}
  \graphicspath{{../pdf/}{../jpeg/}}
  \DeclareGraphicsExtensions{.pdf,.jpeg,.png}
\else
  \usepackage[dvips]{graphicx}
  \graphicspath{{../eps/}}
  \DeclareGraphicsExtensions{.eps}
\fi
\usepackage{amsmath}
\ifCLASSOPTIONcompsoc
\usepackage[tight,normalsize,sf,SF]{subfigure}
\else
\usepackage[tight,footnotesize]{subfigure}
\fi

\usepackage{cite}
\usepackage{url}
\usepackage{listings}
\usepackage{booktabs, multicol, multirow}
\usepackage{amssymb}
\usepackage{pifont}
\usepackage{enumerate}
\usepackage{listings}
\usepackage{color}
\usepackage[dvipsnames]{xcolor}
\usepackage{soul}

\IEEEoverridecommandlockouts

\begin{document}

\title{An Empirical Study of Intel Xeon Phi\thanks{A new version of the work, entitled `Test-Driving Intel Xeon Phi', has been accepted by ICPE 2014, Dublin, Ireland.}}




\author{\IEEEauthorblockN{Jianbin Fang \IEEEauthorrefmark{1},
Ana Lucia Varbanescu \IEEEauthorrefmark{2},
Henk Sips \IEEEauthorrefmark{1}, \\
Lilun Zhang \IEEEauthorrefmark{3}, 
Yonggang Che \IEEEauthorrefmark{3} and 
Chuanfu Xu \IEEEauthorrefmark{3}}

\IEEEauthorblockA{\IEEEauthorrefmark{1}Delft University of Technology, the Netherlands\\ 
Email: \{j.fang, h.j.sips\}@tudelft.nl}
\IEEEauthorblockA{\IEEEauthorrefmark{2}University of Amsterdam, the Netherlands\\ 
Email: \{a.l.varbanescu\}@uva.nl}
\IEEEauthorblockA{\IEEEauthorrefmark{3}National University of Defense Technology, China\\ 
Email: \{llzhang, ygche, xuchuanfu\}@nudt.edu.cn}}

\maketitle

\begin{abstract}
With at least 50 cores, Intel Xeon Phi is a true many-core architecture. Featuring fairly powerful cores, two cache levels, and very fast interconnections, the Xeon Phi can get a theoretical peak of 1000 GFLOPs and over 240 GB/s. These numbers, as well as its flexibility - it can be used both as a coprocessor or as a stand-alone processor - are very tempting for parallel applications looking for new performance records.

In this paper, we present an empirical study of Xeon Phi, stressing its performance limits and relevant performance factors, ultimately aiming to present a simplified view of the machine for regular programmers in search for performance. 

To do so, we have micro-benchmarked the main hardware components of the processor - the cores, the memory hierarchies, the ring interconnect, and the PCIe connection. We show that, in ideal microbenchmarking conditions, the performance that can be achieved is very close to the theoretical peak, as given in the official programmer's guide. We have also identified and quantified several causes for significant performance penalties. Our findings have been captured in four optimization guidelines, and used to build a simplified programmer's view of Xeon Phi, eventually enable the design and prototyping of applications on a functionality-based model of the architecture. 
\end{abstract}

\IEEEpeerreviewmaketitle

\lstset{
language=C,
basicstyle=\scriptsize\sffamily,
frame=single, 
columns=fullflexible,
showstringspaces=false
}

\section{Introduction}
Intel Xeon Phi is the latest high-throughput architecture targeted at high performance computing, and, without a doubt, will be part of the very next generation of supercomputers that will challenge TOP500\footnote{In June 2013, two Xeon Phi supercomputers - \textsc{Tianhe-2} and \textsc{Stampede} - were ranked first and sixth in TOP500 \url{http://www.top500.org.}}. To achieve its high level performance (1000 GFlops), Intel Xeon Phi~\cite{website:intel_mic} uses over 50 cores and 25 MB of on-chip caches. Despite the features it shares with multi-core CPUs and many-core GPUs (vectorization, SIMD/SIMT, high throughput, and high bandwidth)~\cite{citeulike:7370591}, Xeon Phi has a different architecture from all of them~\cite{citeulike:12204757}. For example, overall cache coherency is not available on GPUs, while the ring interconnect is not used on CPUs and GPUs.

For advanced users - like most high performance computing (HPC) programmers and compiler developers are - it is essential to understand this architecture in detail, as the achieved performance depends on each of these details. For example, knowing the requirements for density and placement of threads per cores, the optimal utilization of the core interconnections, or the difference in latency between the different types of memories on chip are non-trivial details that, when properly exploited, can lead to significant performance gains. 

On the other hand, normal, productivity-oriented programmers need a simplified view of the Xeon Phi machine, which captures only the key performance features, easing the algorithmic design and high-level programming without significant performance losses.

Therefore, in this paper, we focus on answering three questions: 
(1) \textbf{what are the high-impact performance factors for each architectural component of Xeon Phi?},  
(2) \textbf{are the theoretical performance numbers published for Xeon Phi achievable in practice?}, and   
(3) \textbf{can we build a simplified, functionality-based model of Xeon Phi to simplify high-level programming?}.
 
As empirical evaluation based on benchmarking is already recognized as a suitable solution for achieving this level of understanding~\cite{citeulike:9219351}, our work is based on a comprehensive suite of microbenchmarks. 
Specifically, we use dedicated microbenchmarks to measure the performance of the key architectural features of Xeon Phi: the processing cores, the memory hierarchies, the ring interconnect, and the PCIe connection. To capture the capabilities of the different types of components, we use both latency-oriented metrics (e.g., seconds, cycles), and throughput-oriented metrics (e.g., GFlops, GB/s), depending on the capability at test. Our methodology, together with the benchmarks, are presented in more detail in Section~\ref{sec:methodology}. We note that the results obtained with our microbenchmarks can already be used to gain a deeper understanding of the hardware behavior (at times, complementary to the official specification), but will further prove very useful for establishing empirical upper bounds of the achievable performance for applications, and for building hardware performance models.




The main contributions of our work are as follows:
\begin{itemize}

\item We present a first comprehensive evaluation of Xeon Phi (Section~\ref{sec:benchmarking}), obtaining interesting numerical results for the capabilities of its cores, memories, on-chip and off-chip (i.e., PCIe) interconnect.


\item We synthesize four essential platform-centric performance requirements, aimed at easing the development and tuning of applications (Section~\ref{sec:comparison}). Although our claims are not contradictory to Intel's documentation, we provide much more practical information.

\item We propose a simplified model of Xeon Phi, which strips off the performance irrelevant architectural details, presenting the programmers with a simple, functionality-based view of the machine to be programmed with no significant performance loss (Section~\ref{sec:model}). 

\end{itemize}

\section{Benchmarking Intel Xeon Phi} \label{sec:bm_background}
In this section, we introduce the Intel Xeon Phi architecture - with its novel features and typical programming models, and we present our benchmarking methodology. 
 
\subsection{The Architecture}
\begin{figure}[!th]
\centering
\includegraphics[width=0.35\textwidth]{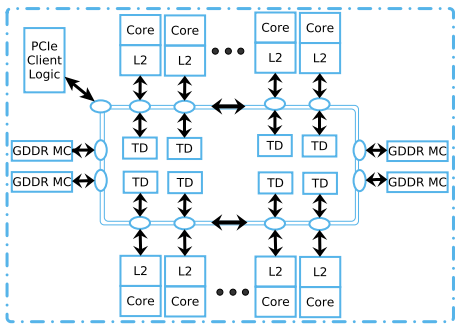}
\caption{The Intel Xeon Phi Architecture.}
\label{fig:mic_architecture}
\end{figure}

Intel Xeon Phi has over 50 cores (the version used in this paper belongs to the 5100 series and has 60 cores) connected by a high-performance on-die bidirectional interconnect (shown in Figure~\ref{fig:mic_architecture}). In addition to these cores, there are 16 memory channels (supported by memory controllers) delivering up to 5.0 GT/s~\cite{citeulike:12204757}. When working as an accelerator, Phi can be connected to a host (i.e., a device that manages it) through a PCI Express (PCIe) system interface - similar to GPU-like accelerators. Different from GPUs, a dedicated embedded Linux $\mu$OS (version: \texttt{2.6.38.8}) runs on the platform.

Each core contains a 512-bit wide vector unit (VPU) with vector register files (32 registers per thread context). 
Each core has a 32KB L1 data cache, a 32KB L1 instruction cache, and a core-private 512KB unified L2 cache. In total, a 60-core machine has a total of 30MB of L2 cache on the die. The L2 caches are kept fully coherent by the hardware, using DTDs (distributed tag directories), which are referenced after an L2 cache miss. 
Note that the tag directory is not centralized, but split up into 64 distributed tag directories (DTDs), each getting an equal portion of the address space and being responsible for maintaining it globally coherent. 
Another special feature of Xeon Phi is the fast bidirectional ring interconnect. All connected entities use the ring for communication purposes, using special controllers called \textit{ring stops} to insert requests and receive responses on the ring. 

The novelties of the Xeon Phi architecture relate to five components : (1) the vector processing cores, (2) the on-chip memory, (3) the off-chip memory, (4) the ring interconnect, and (5) the PCIe connection. As these are the features that differ, in one way or another, from a typical CPU - vectors are wider, there are many more cores, cache coherency and shared memory are provided with low penalty for 60 or more cores, and a ring interconnect holds tens of agents that can interchange messages/packets concurrently -, we focus our benchmarking efforts on these features. 

We also note that the similarities between Phi and a regular multi-core CPU allow us  to adapt existing CPU benchmarks to the requirements of Xeon Phi. In most cases, we use such "refurbished" solutions, that prove to serve our purposes well. 


\subsection{Programming}

In terms of usability, there are two ways an application can use Intel Xeon Phi: (1) in \textit{offload mode} - the main application is running on the host, and it only offloads selected (highly parallel, computationally intensive) work to the coprocessor, or (2) in  \textit{native mode} - the application runs independently, on the Xeon Phi only, and can communicate with the main processor or other coprocessors~\cite{citeulike:12214076} through the system bus. In this work, we benchmark Xeon Phi in both modes. 

Finally, to program applications on Xeon Phi, users need to capture both functionality and parallelism. 
Being an \texttt{x86} SMP-on-a-chip architecture, Xeon Phi offers the full capability to use the same tools, programming languages, and programming models as a regular Intel Xeon processor. Specifically, tools like Pthreads~\cite{citeulike:12606769}, OpenMP~\cite{citeulike:12606776}, Intel Cilk Plus~\cite{citeulike:4457581}, and OpenCL~\cite{stone-2010} are readily available. Given the large number of cores on the platform, a dedicated MPI version is also available. 
In this work, all the experiments we present are programmed useing C, OpenMP, and/or Pthreads; we also use Intel's \texttt{icc} compiler (\texttt{V13.1.1.163}). 


\subsection{Benchmarking Methodology}
\label{sec:methodology}

The goal of our benchmarking is two-fold: to show how the special capabilities of Xeon Phi can and should be measured, and to quantify the performance of this novel many-core architecture. Therefore, we choose for a microbenchmarking approach: we measure each capability in isolation, under variable loads, and we quantify its performance in terms of both latency-oriented and throughput-oriented metrics. As Xeon Phi is intended for high-performance computing, low latency and high throughput are essential to understand the performance achievable for different applications~\cite{citeulike:5836618,citeulike:6778359}.    

Simply put, \textit{latency} is the time required to perform an operation and produce a result. As lower latency means better performance, microbenchmarking aims to measure the \emph{minimum achievable latency} for various types of operations; typically, the benchmarked values are expected to be slightly higher than the theoretical ones. As latency measurement focuses on a single action from its beginning to its end, one needs to isolate the operation to be measured and use a highly accurate, non-intrusive timing method. However, this is difficult to achieve on modern processors, where operation latency is in the order of (tens of) nanoseconds. Therefore, an alternative solution is used: we measure a long enough sequence of operations with an accurate timer, and estimate latency per operation by dividing the measured time by the number of operations. 

In this paper, latency measurements are done with a single thread (for individual operations) or two threads (for transfer operations) with Pthreads. All latency benchmarks are written in C (with inline assembly) and compiled with no optimizations (i.e., we force the compiler to ignore any opportunities to eliminate or reorder operations, which would alter the intended instruction mix). 



\textit{Throughput} is the number of (a type of) operations executed in a given unit of time. As higher throughput means better performance, microbenchmarking focuses on measuring the \emph{maximum achievable throughput} for different operations, under different loads; typically, the benchmarked throughput values are slightly lower than the theoretical ones. Thus, 
to measure maximum throughput, the main challenge is to build the workload such that the resource that is being evaluated is fully utilized. 
For example, when measuring computational throughput, enough threads should be used to fully utilize the cores, while when measuring memory bandwidth, the workload needs to have sufficient threads to generate enough memory requests.

For all the throughput measurements in this paper, our multi-threaded workloads are written in C using OpenMP. This allows us to easily adjust the number of threads to experiment with different loads. 




\lstdefinestyle{customc}{
  belowcaptionskip=1\baselineskip,
  breaklines=true,
  frame=L,
  xleftmargin=\parindent,
  language=C,
  showstringspaces=false,
  basicstyle=\footnotesize\ttfamily,
  keywordstyle=\bfseries\color{green!40!black},
  commentstyle=\itshape\color{purple!40!black},
  identifierstyle=\color{black},
  stringstyle=\color{orange},
  morekeywords={\#, pragma, omp, parallel},
}

\lstdefinestyle{customasm}{
  belowcaptionskip=1\baselineskip,
  frame=L,
  xleftmargin=\parindent,
  language=[x86masm]Assembler,
  basicstyle=\footnotesize\ttfamily,
  commentstyle=\itshape\color{purple!40!black},
}

\lstset{escapechar=@,style=customc}

%
%


\section{Empirical Evaluation} \label{sec:benchmarking}

In the following sections, we present in detail the microbenchmarks we have used and their results for each of the components: (1) the vector processing cores, (2) the on-chip memory, (3) the off-chip memory, (4) the ring interconnect, and (5) the PCIe connection.


\subsection{Vector Processing Cores} \label{subsec:processing_cores}
We evaluate the vector processing cores in terms of both instruction latency and throughput. For latency, we a method similar to those proposed by Agner Frog~\cite{citeulike:12589493} and Torbjorn Granlund~\cite{citeulike:12589489}: we measure instruction latency by running a (long enough) sequence of dependent instructions (i.e., a list of instructions that, being dependent on each other, are forced to be executed sequentially - \emph{an instruction stream}).

The same papers propose a similar approach to measure throughput in terms of \textit{instruction per cycle (IPC)}. However, we argue that a measurement that uses all processing cores together, and not in isolation, is more realistic for programmers. Thus, we use the \texttt{flops} microbenchmark from the \texttt{shoc-mic}\footnote{\url{https://github.com/vetter/shoc-mic}} suite to explore the factors for reaching the theoretical maximum throughput on Xeon Phi (Section~\ref{subsec:vec_thr}). 
Finally, we have designed a microbenchmark to show the performance differences between single-precision and double-precision operations, knowing that Xeon Phi has an EMU (extended math unit) only for single-precision operations. 

\subsubsection{Vector Instruction Latency}
Xeon Phi introduces 177 vector instructions~\cite{citeulike:12589702}. We roughly divide these instructions into five classes~\footnote{Note that we choose not to measure the latency of memory access instructions because the latency results are highly dependent on the data location(s).}: mask instructions, arithmetic (logic) instructions, conversion instructions, permutation instructions, and extended mathematical instructions. 

The benchmark for measuring the latency of vector instructions is measuring the execution time of a sequence of 100 vector operations using the same format: $zmm1 = op(zmm1, zmm2)$, where $zmm1$ and $zmm2$ represent two vectors and $op$ is the instruction being measured. By making $zmm1$ be both a source operand and the destination operand, we ensure the instruction dependency - i.e., the current operation will depend on the result of the previous one. 
 
For special classes of instructions - such as the \texttt{conversion} instructions \texttt{vcvtps2pd} and \texttt{vcvtpd2ps} - we have to measure the latency of the conversion pair ($zmm2 = op12(zmm1); zmm1 = op21(zmm2)$) in order to guarantee the dependency between contiguous instructions (i.e., it is not possible to write the result of the conversion in the same source operand, due to type incompatibility). Similarly, we measure the latency of extended mathematical instructions such as \texttt{vexp223ps} and \texttt{vlog2ps} in pairs, to avoid overflow (e.g., when using 100 successive \texttt{exp()}'s). 

The interesting results for vector instructions latency are presented in Table~\ref{tbl:ins_lat}. We see that the latency of a \texttt{mask} instruction is 2 cycles (and not 1 cycle, as reported the official documentation~\cite{citeulike:12204757}). 
The \texttt{arithmetic} vector instructions take 4 cycles, while \texttt{conversion} instructions take 5 cycles. The most expensive instructions are the \texttt{permutation} and \texttt{extended mathematical} instructions, with 6 cycles. With these latency numbers, we know how many threads or instruction streams we need to hide the latency. 

\begin{table}[!h]
\caption{The latency of vector instructions on Xeon Phi (in cycles)}
\begin{center}
\begin{tabular}{|c|l|c|}
\hline \hline
\textbf{Instruction} & \textbf{Category} & \textbf{Latency} \\ \hline
\begin{tabular}[c]{@{}c@{}}kand, kor,\\knot, kxor\end{tabular}  & mask instructions & 2 \\ \hline
\begin{tabular}[c]{@{}c@{}}vaddpd, vfmadd213pd,\\vmulpd, vsubpd\end{tabular} & arithmetic instructions & 4 \\ \hline
\begin{tabular}[c]{@{}c@{}}vcvtdq2pd, vcvtfxpntdq2ps,\\vcvtfxpntps2dq, vcvtps2pd\end{tabular} & convert instructions & 5 \\ \hline
vpermd, vpermf32x4 & permutation instructions & 6 \\ \hline
\begin{tabular}[c]{@{}c@{}}vexp223ps, vlog2ps,\\vrcp23ps, vrsqrt23ps\end{tabular}  & \begin{tabular}[c]{@{}c@{}}extended \\mathematical instructions\end{tabular} & 6 \\ \hline \hline
\end{tabular}
\end{center}
\label{tbl:ins_lat}
\end{table}


\subsubsection{Vector Instruction Throughput} \label{subsec:vec_thr}


The Xeon Phi 5100 has 60 cores working at 1.05 GHz, and each core can process 8 double-precision data elements at a time, with maximum 2 operations (\texttt{multiply-add} or \texttt{mad}) per cycle in each lane (i.e., a vector element). 
Therefore, the theoretical instruction throughput is 1008~GFlops (approximately 1~TFlop). But \textbf{is this 1~TFlop performance actually achievable?} 
To measure the instruction throughput, we run 1, 2, 4 threads on a core (60, 120, and 240 threads in total).  
During measurement, each thread performs one or two instruction streams for a fixed number of iterations: $b_{i+1} = b_{i}\ op\ a$, where $i$ represents the iteration, $a$ is a constant, and $b$ serves as an operand and the destination. The loop was fully unrolled to avoid branch overheads.
The microbenchmark is vectorized using explicit intrinsics, to ensure a 100\% vector usage.

\begin{figure}[!t]
\centering
\includegraphics[width=0.35\textwidth]{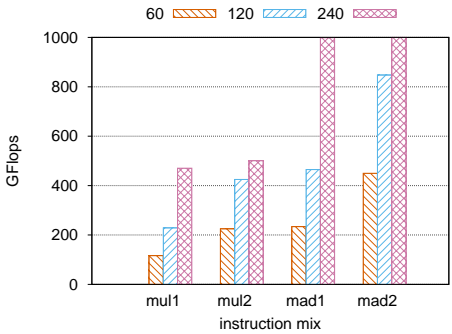}
\caption{Arithmetic throughput using different numbers of threads (60, 120, 240), different instruction mixes (\texttt{mul} versus \texttt{mad}), and issue widths (using one and two independent instruction streams).}
\label{fig:arithmetic_throughput}
\end{figure}

The results are shown in Figure~\ref{fig:arithmetic_throughput}. We note that the peak instruction throughput - i.e., one vector instruction per cycle (1TFlops in total) - can be achieved when using 240 threads and the multiply-add instruction. As expected, the \texttt{mad} throughput is twice larger than the \texttt{mul} throughput. 
Further, two more observations can be added. 
First, when using 60 threads (one thread per core), the instruction throughput is low compared with the cases when using 120 or 240 threads. This is due to the fact that it is not possible to issue instructions from the same thread context in back-to-back cycles~\cite{citeulike:12204757}. Thus, programmers need to run at least two threads on each core to be able to fully utilize the hardware resources. 
Second, when a thread is using only one instruction stream, we have to use 4 threads per core (240 threads in total) to achieve the peak instruction throughput. This is because the latency of an arithmetic instruction is 4 cycles (Table~\ref{tbl:ins_lat}), and we need no less than four threads to totally hide this latency (i.e., fill the pipeline bubbles~\cite{citeulike:12607692}). 
To comply, programmers need to either use 4 threads per core or have more independent instruction streams. 

To summarize, for a given instruction mix (\texttt{mul} or \texttt{mad}), the achievable instruction throughput depends not only on the number of cores and threads, but also on the issue width (i.e., the number of independent instruction streams). 


\textbf{Extended Math Unit (EMU):}  Extended math functions such as \texttt{exp}, \texttt{log}, \texttt{sin}, and \texttt{cos} are often used in scientific computing. The VPU of Xeon Phi includes an EMU (extended math unit) which can perform base-2 exponential, base-2 logarithm, reciprocal, and reciprocal square toot of \textit{single precision} operands~\cite{citeulike:12204757}. Their instruction latencies are shown in Table~\ref{tbl:ins_lat}. For the double precision versions, Xeon Phi does not have the corresponding execution unit. 
To determine the performance gain that the EMU brings to execute these computations on Xeon Phi, we compare the performance of the EMU-based single-precision operations with that of the non-EMU double-precision operations. We take as an example the exponential calculation: $B[k]=exp(A[k])$ (where A and B are arrays of 1024 floating point elements, single or double precision). We run this computation 100,000 times on a single thread and measure its execution time. We use four different versions of the $exp$ operation: the intrinsics implementations - \texttt{svml} (for base-$e$) and \texttt{svml2} (for base-$2$), and the C library implementations - \texttt{clib} and \texttt{clib2}, respectively. 

\begin{figure}[!t]
\centering
\includegraphics[width=0.35\textwidth]{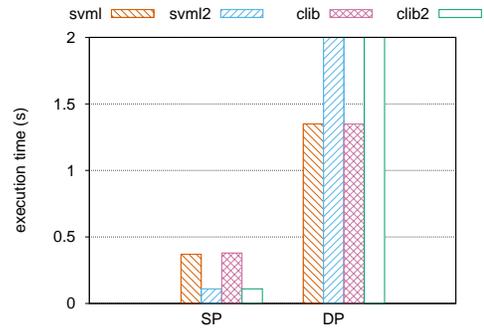}
\caption{A comparison of the \texttt{exp} performance: single versus double precision data, using four different versions of functions for calculating exponentials.}
\label{fig:arithmetic_throughput_exp}
\end{figure}

In Figure~\ref{fig:arithmetic_throughput_exp}, we show the performance results for all four implementations, for both single and double precision. We first note that there are very small performance differences between the equivalent versions. Thus, programmers can use the C library implementations for simplicity. 
Further, using the base-$2$ implementations (svml2 and clib2) can significantly decrease the execution time for the single-precision \texttt{exp} (around 4 times faster), whereas the performance decreases by around 30\% on the double-precision data. Additionally, compared with the double-precision version of \texttt{exp}, using single-precision data elements perform 5 times faster. Therefore, single-precision computations are recommended (for math-intensive kernels) when accuracy requirements can still be met.



\subsection{Memory Latency}
Available benchmarks, such as \texttt{BenchIT}~\cite{website:benchit_benchmark} and \texttt{lmbench}~\cite{citeulike:12586247} use \textit{pointer-chasing} to measure the on-chip and off-chip memory access latency. This approach has the advantage of not only determining the latency itself, but also exposing the differences between consecutive layers of a memory hierarchy (i.e., different layers of caches and main memory will have significantly different latencies). 
Thus, we use a similar approach to measure the latency for a Xeon Phi core (i.e., the latency for accessing local caches and main memory - see Section~\ref{subsec:cache_property}). 

When more than two cores communicate, measuring latency is complicated. For this, Daniel Molka et al. proposed an approach to quantify cache-coherency effects~\cite{citeulike:8778343}. In our work, we adapt this approach to Xeon Phi using the correct memory fences and cache flushing instructions~\footnote{Since Xeon Phi has no \texttt{mfence} or \texttt{clflush}, we need to change the benchmark by searching and replacing them with equivalent instructions.}.

\subsubsection{Access Latency on a Single Core} \label{subsec:cache_property}
To reveal the local access latency, we use a \textit{pointer-chasing} benchmark similar to those used by \texttt{BenchIT} and \texttt{lmbench}. Essentially, the application traverses an array $A$ of size $S$ by running $k=A[k]$ in a 
fully unrolled loop. The array is initialized with a $stride$, i.e., $A[k]=(k+stride)\%S$. By measuring the execution time of the traversal, we can easily obtain an estimate of the average execution time for one iteration. This time is dominated by the latency of the memory access. The traversal is done in one thread and utilizes only one core. Therefore, the memory properties obtained here are local and belong to one core.

\begin{figure}[!h]
\centering
\includegraphics[width=0.35\textwidth]{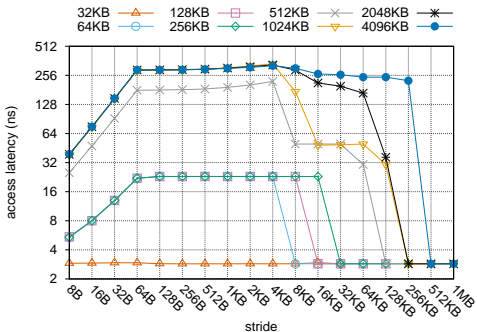}
\caption{Average memory latency when changing strides and datasets. The x-axis is logarithmic and it represents the pointer chasing stride.}
\label{fig:stride_tot}
\end{figure}

The results are shown in Figure~\ref{fig:stride_tot}. We see that the Xeon Phi has two levels of data caches (L1 and L2). The L1 data cache is 32KB, while the L2 data caches should be smaller than 512KB. Furthermore, the accessing latency of L1 and L2 data caches is around 2.87 ns (3 cycles)  and 22.98 ns (24 cycles), respectively. With a stride of 64 bytes, Xeon Phi takes 287.51 $\sim$ 291.18 ns (302 $\sim$ 306 cycles) to finish a data access in the main memory (when the dataset is larger than 512KB). 
We note that when traversing the array in a larger stride (e.g., 4KB), the latency of accessing data in off-chip memory is slightly larger. This is because the contiguous memory accesses fall into different pages. 



When the data elements fit inside the L2 cache (i.e., the array size is smaller than 256 KB), we can take a closer look at the L1 cache. We see that when the array is no larger than 32KB (L1 size), the access time stays stable, since all the data elements can be stored in the L1 cache. When the input array is larger than 32KB, the access time increases over strides. This trend continues until the stride is 64 bytes, which shows that the threads operate the data in a batch manner, i.e., a cache-line. This result indicates that, for achieving good performance, applications need to access the memory in contiguous batches (as expected). 
When the stride is larger than 64 bytes, the accesses will experience a L1 cache miss per access, with the latency staying stable. Note that more information about cache associativity can also be seen in Figure~\ref{fig:stride_tot} (see~\cite{citeulike:9482128} for the calculation approach).

\subsubsection{Remote Cache Latency} \label{subsec:remote_lat}
We have illustrated our measurements and results for memory latency on a single core in Section~\ref{subsec:cache_property}. However, for making right parallelization decisions on such a many-core platform, it is important to have a good understanding of the latency of non-local memory accesses. Thus, in this section, we focus on measuring remote cache latency.

For these measurements, we use an approach based on that proposed by Daniel Molka~\cite{citeulike:8778343}. Our setup is built as follows: prior to the measurement, the to-be-transferred cache-lines are placed in different locations (cores) and in a certain coherency state (\texttt{modified}, \texttt{exclusive}, or \texttt{shared}). In each measurement, we use two threads (T0, T1), with T0 pinned to Core 0 and T1 pinned on another core (Core $X$). The latency measurement always runs on Core 0, transferring a predefined number of cache lines from Core $X$ to Core 0. 

\begin{figure}[!h]
\centering
\subfigure[Modified]{\label{fig:latency_m}\includegraphics[width=0.35\textwidth]{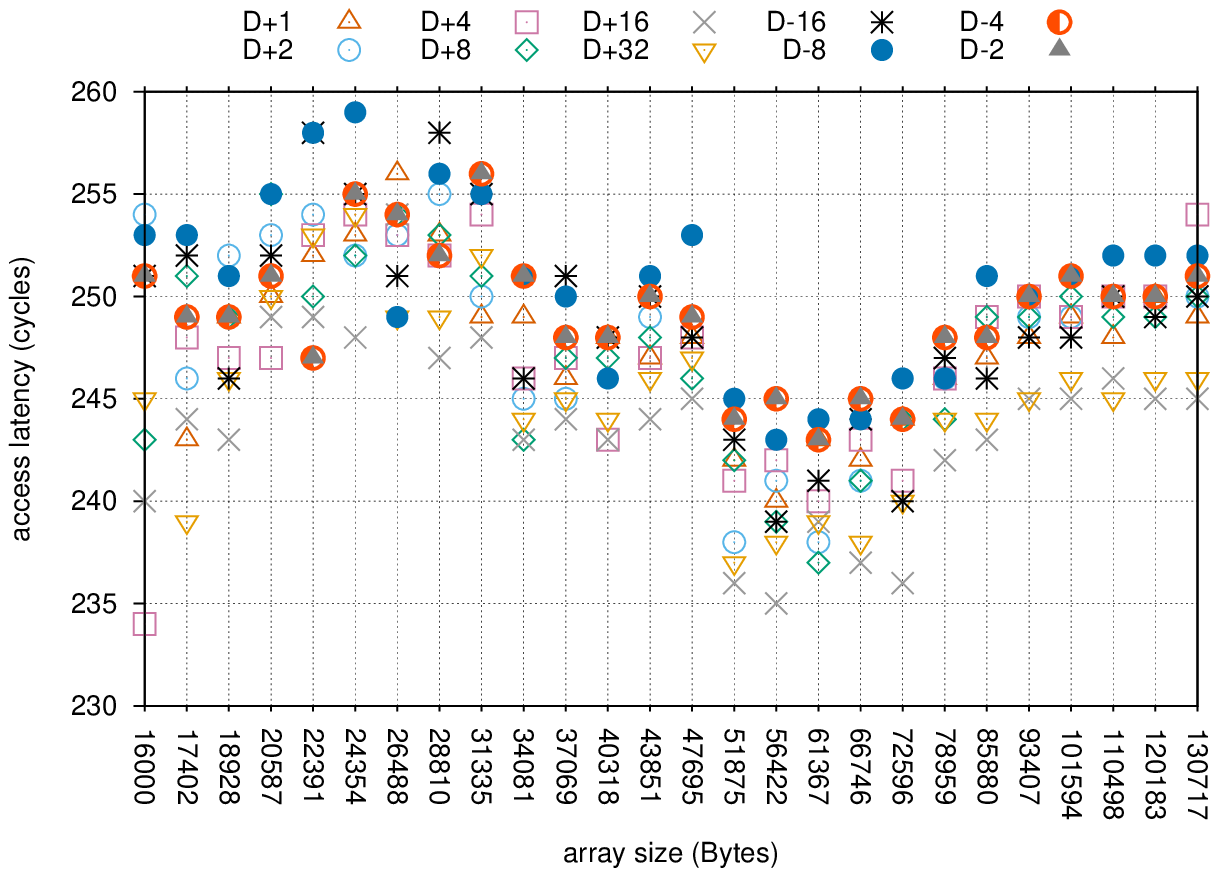}}
\subfigure[Overall]{\label{fig:latency_tot}\includegraphics[width=0.35\textwidth]{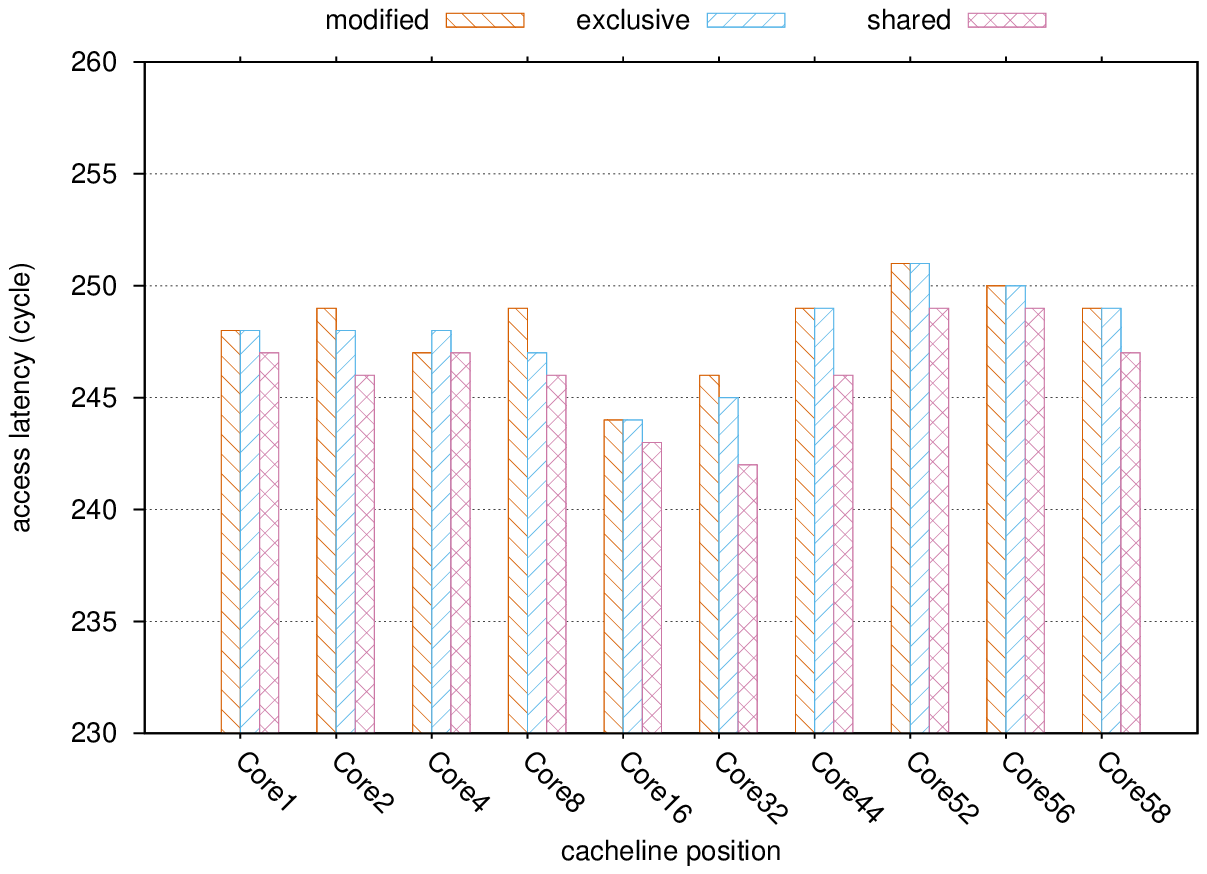}}
\caption{Read latencies of Core 0 accessing the cache lines on Core 1 (D+1), Core 2 (D+2), Core 4 (D+4), Core 8 (D+8), Core 16 (D+16), Core 32 (D+32), Core 44 (D-16), Core 52 (D-8), Core 56 (D-4),and Core 58 (D-2).}
\label{fig:latency_remot}
\end{figure}

Figure~\ref{fig:latency_remot} shows our results for remote cache accesses latency on Xeon Phi. In Figure~\ref{fig:latency_m}, we see that when the cache line is in \texttt{modified} state, the overall latency of remote access averages around 250 cycles (roughly matching the results by Garea~\cite{id_ramos_hpdc}), which is much larger than the local cache access latency (by an order of magnitude) but still smaller than the off-chip memory access latency (by 17\%). By getting the \textit{median} value of all the input data sets (up to 128 KB), we get the overall remote latency shown in Figure~\ref{fig:latency_tot}. 
We note no relationship between the remote access latency and the cache-line states, except that accessing remote \texttt{shared} cachelines takes a few less cycles. This is because in whichever state a cacheline is, when a core accesses it, a transfer is needed from a remote core. 
Furthermore, Xeon Phi adopts the \texttt{MOESI} cache coherence protocol~\cite{citeulike:12204757} to share a cacheline before writing it back, and thus Figure~\ref{fig:latency_tot} shows no penalty of writing data back. In~\cite{myself_bm_phi}, our experiments have shown that there is a relation between the latency and the core distances on an older version of the Xeon Phi (namely, 31S1P), but this effect seems to have disappeared on the newer Xeon Phi 5110.



To summarize, as expected, best performance is achieved when each core keeps data in its own local cache, and avoids remote cache and off-chip memory accesses as much as possible. 
With regard to cache capacity, the Xeon Phi uses a private LLC (last-level cache) of size 512KB~\cite{citeulike:12250643}. Specifically, if no cores share any data or code, the effective total L2 size of the chip is 30 MB (with 60 cores). Whereas, if every core shares exactly the same data and code in perfect synchronization, then the effective total L2 size of the chip is 512 KB. To save the scarce on-chip resource, programmers need to take care of the data arrangement when partitioning tasks, and put the tasks with data sharing onto the same core(s).

\subsection{Memory Bandwidth}
McCalphin's stream benchmark~\cite{website:stream_benchmark} includes a memory bandwidth benchmark and presents results for a large number of high-end systems. However, his solution is based on a combination of both read and write operations. In this paper, we want to separate \textit{reads} and \textit{writes}, and make a bandwidth estimation for more real scenarios. 
In \texttt{BenchIT}, Daniel Molka et al. presents a solution to measure bandwidth in a similar way with that of latency measurement (see Section~\ref{subsec:remote_lat}). His microbenchmark requires compiler optimizations to be disabled (i.e., the code should be compiled with the \texttt{-O0} option), thus disabling the software prefetching on Xeon Phi. As a result, this measurement will underestimate bandwidth. 
\texttt{lmbench} attempts to measure system's ability to transfer data between processor, cache, memory, disk, and network. However, \texttt{lmbench} supports parallelism in terms of \texttt{Unix processes}, which have to use special inter-process communication methods instead of data sharing within a single data space. 
In this section, we present our own \texttt{OpenMP} implementation of a memory bandwidth microbenchmark, an adaptation of the \texttt{lmbench} version to the Xeon Phi architecture. 

\subsubsection{Off-Chip Memory Bandwidth} \label{subsec:off_chip_mem_bw}
The Xeon Phi used in this work has 16 memory channels, each 32-bits wide. At up to 5.0 GT/s transfer speed~\footnote{GT/s stands for Giga Transfers per second.}, 
it provides a theoretical bandwidth of 320 GB/s.
But \textbf{is this theoretical bandwidth really achievable in real cases?} 
To answer this question, we use separate benchmarks to measure the memory bandwidth for both \texttt{read} and \texttt{write} operations. The \texttt{read} benchmark reads data from an array $A$ ($b=b+A[k]$). The \texttt{write} benchmark writes a constant value into an array $A$ ($A[k]=C$). Note that $A$ needs to be large enough (e.g., 1 GB) such that it cannot fit in the on-chip memory. 
To avoid the impact of "cold" TLBs, we start with two "warm-up" iterations of the benchmarks, before we measure a third one; the caches are flushed between iterations. We use different numbers of running threads - from 1 to 240.

\begin{figure}[!t]
\centering
\subfigure[read]{\label{fig:read}\includegraphics[width=0.35\textwidth]{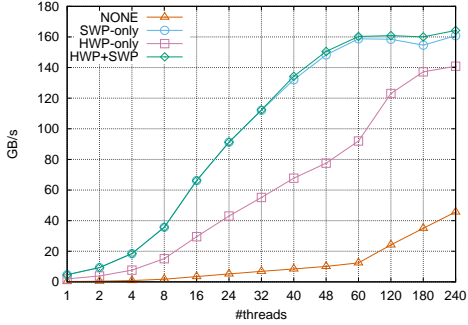}}
\subfigure[write]{\label{fig:write}\includegraphics[width=0.35\textwidth]{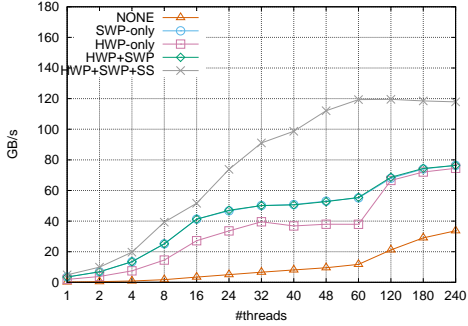}}
\caption{Read and write memory bandwidth.}
\label{fig:off_chip_mem_bw}
\end{figure}

Our results are shown in Figure~\ref{fig:off_chip_mem_bw} (HWP+SWP) (we plot the median value of ten runs of the benchmarks). Overall, we see that the maximum bandwidth for both \texttt{read} and \texttt{write} is far below the theoretical peak of 320 GB/s. Moreover, both the \texttt{read} and \texttt{write} memory bandwidth increases over the number of threads - which happens because when using more threads, we can generate more requests to memory controllers, thus making the interconnect and memory channels busier. Thus, if aiming to achieve high memory bandwidth, programmers need to launch enough threads to saturate the interconnect and the memory channels. 
Figure~\ref{fig:read} also shows that the \texttt{read} bandwidth peaks at 164 GB/s, achievable with using 60 threads or more (pinning at least one thread to a core). 
However, we can obtain the maximum \texttt{write} bandwidth (76 GB/s, as seen in Figure~\ref{fig:write}) only when using 240 threads. 
In general, the \texttt{write} bandwidth is around half of the \texttt{read} bandwidth. This happens because Xeon Phi implements a write-allocate cache policy and the original content has to be loaded into caches before we overwrite it completely. 
To avoid the memory bandwidth waste, programmer can use \textit{streaming stores}~\footnote{Streaming stores do not require a prior cache line read for ownership (RFO) but write to memory ''directly''.} on Xeon Phi~\cite{citeulike:12533940}. 
We see that using \textit{streaming store} instructions speeds-up write operations up to 1.7 times (Figure~\ref{fig:write}:HWP+SWP+SS), with memory write bandwidth now peaking at 120 GB/s.
Thus, programmers must consider using \textit{streaming stores} to optimize the memory bandwidth. 

\textbf{Prefetch Effects:} Xeon Phi supports both hardware prefetching (HWP) and software prefetching (SWP). The L2 cache has a streaming hardware prefetcher that can selectively prefetch code, read, and RFO (Read-For-Ownership) cachelines into the L2 cache~\cite{citeulike:12204757}. Figure~\ref{fig:off_chip_mem_bw} shows the memory bandwidth of four different configurations: no prefetching, HWP or SWP only, or both. 
Overall, the figure shows that memory bandwidth increases over the number of threads. When disabling both HWP and SWP, the memory bandwidth is low (45 GB/s for reading and 33 GB/s for writing). With only SWP, we already achieve similar memory bandwidth to that achieved when enabling both of them. This similarity indicates that the hardware prefetcher will not kick in when software prefetching performs well. 
Furthermore, enabling only HWP only delivers about half of the bandwidth achieved when enabling only SWP (the bandwidth is roughly 1.9$\times$ smaller, on average). 

\begin{figure}[!t]
\centering
\includegraphics[width=0.35\textwidth]{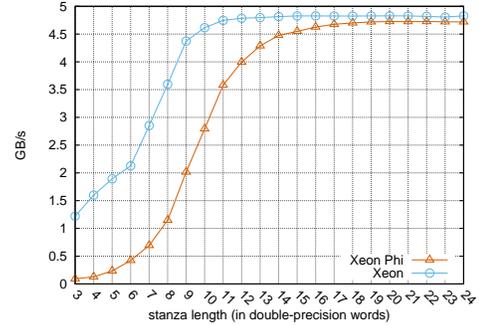}
\caption{Performance of STriad on the Xeon Phi (the x-axis is in log scale and the results on Xeon are normalized to those on Xeon Phi).}
\label{fig:prefetch_stanza}
\end{figure}

To further evaluate the efficiency of prefetching on Xeon Phi, we also use the Stanza Triad (\texttt{STriad})~\cite{citeulike:12272768} benchmark with a single thread. 
\texttt{STriad} works by performing a DAXPY (Triad) inner loop for a length $L$ stanza, then jumps over $k$ elements, and continues with the next $L$ elements, until reaching the end of the array. 
We set the total array size to 128 MB, 
and set $k$ to 2048 double-precision words. 
For each stanza, we ran the experiment 10 times, with the L2 cache flushed each time, and we calculate median value of the 10 runs to get the memory bandwidth for each stanza length. 
Figure~\ref{fig:prefetch_stanza} shows the results of the STriad experiments on both Xeon Phi and a regular Xeon processor (Intel Xeon E5-2620). We see an increase in memory bandwidth over stanza length $L$, and we note it eventually approaches a peak of 4.7 GB/s (note that this is achieved per core). Further, we see the transition point (from the bandwidth-increasing state to the bandwidth-stable state) 
appears earlier on Xeon than on Xeon Phi. 
Therefore, we conclude that non-contiguous access to memory is detrimental to memory bandwidth efficiency, with Xeon Phi showing more restrictions on the stanza length when prefetching data than the regular Xeons.
To comply, programmers have to create the longest possible stanzas of contiguous memory accesses, improving prefetching and memory bandwidth. This is not unexpected, but the impact of not taking into account this requirement is larger than expected. 


\textbf{ECC Effects: }
The Xeon Phi coprocessor supports ECC (Error Correction Code) to avoid software errors caused by naturally occurring radiation. Enabling ECC adds reliability, but it also introduces extra overhead to check for errors. We examined the bandwidth differences with and without disabling ECC. With {ECC} disabled, we noticed a 20\% to 27\% bandwidth increase~\cite{myself_bm_phi}. Note that all the experiments in this paper are performed with ECC enabled.  


\subsubsection{Aggregated On-Chip Memory Bandwidth}
The available on-chip memory bandwidth is always essential in performance tuning and analysis. So, \textbf{how large is the on-chip memory bandwidth that can be achieved?} To answer this question, we measure the cache bandwidth on a single core~\footnote{Note that we choose not to measure the inter-core communication bandwidth because we assume that cache-line transfers occur rather scattered, and not in a large volume. Thus, the measurement of inter-core (remote) access latency is of greater use.} and calculate the \textit{aggregated} cache bandwidth by multiplying it with the number of cores. We first use a set of \texttt{vmovapd} instructions to measure the native \texttt{read} or \texttt{write} bandwidth. Our results show that the L1 access (read or write) throughput is 64 bytes per cycle. Thus, the aggregated L1 bandwidth is 4032 GB/s for \texttt{read} or \texttt{write}. Then we measure the maximum achieved bandwidth from programmers' point of view for \texttt{scale1} ($O[i]=a \times A[i]$), \texttt{scale2} ($O[i]=a \times O[i]$), \texttt{saxpy1} ($O[i]=a \times A[i]+B[i]$), and \texttt{saxpy2} ($O[i]=a \times A[i]+O[i]$) operations. To avoid overheads from the high-level code, we use intrinsics in the kernel code. We also disable the software prefetching due to the fact that the data is located in caches after warming up. 



\begin{figure}[!t]
\centering
\subfigure[scale1 bandwidth.]{\label{fig:scale1}\includegraphics[width=0.235\textwidth]{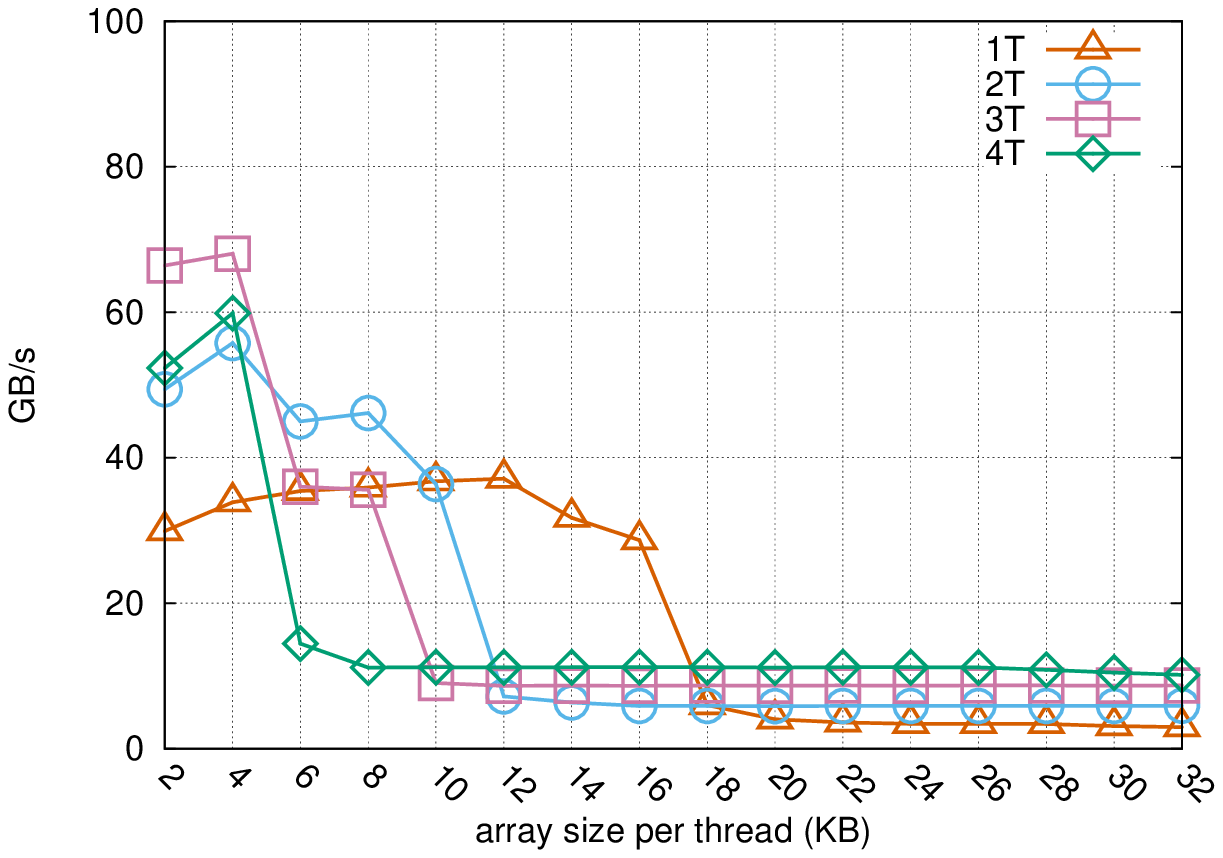}}
\subfigure[scale2 bandwidth.]{\label{fig:scale2}\includegraphics[width=0.235\textwidth]{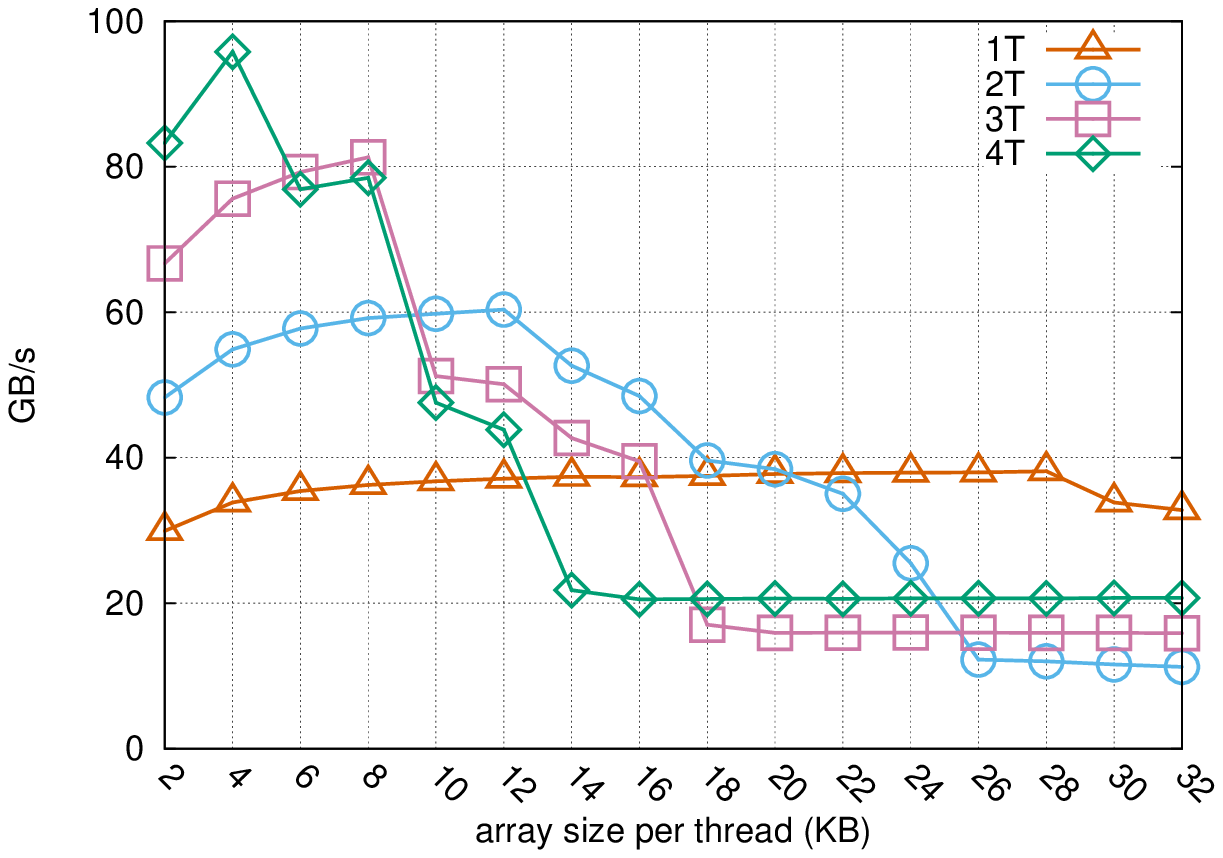}}
\subfigure[saxpy1 bandwidth.]{\label{fig:saxpy1}\includegraphics[width=0.235\textwidth]{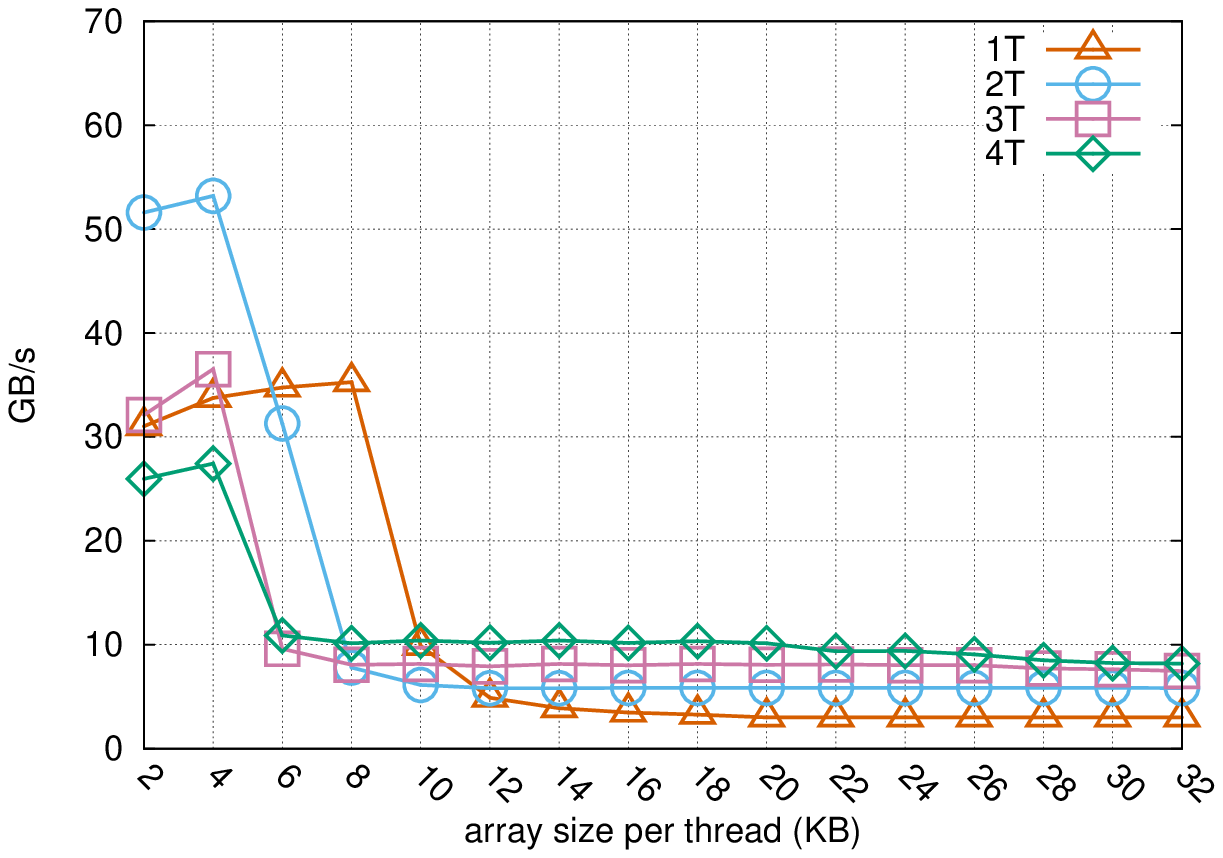}}
\subfigure[saxpy2 bandwidth.]{\label{fig:saxpy2}\includegraphics[width=0.235\textwidth]{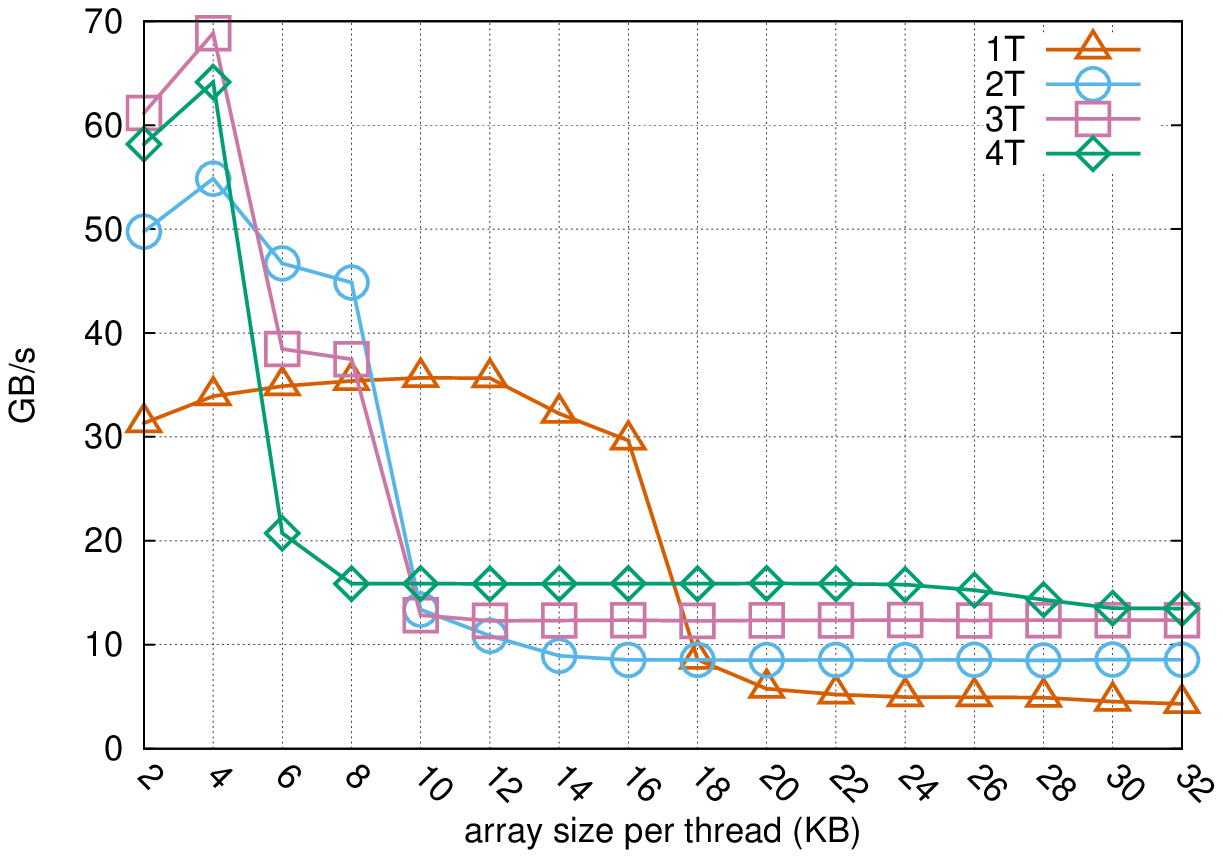}}
\caption{Cache bandwidth on a single core.}
\label{fig:cache_bw}
\end{figure}

The results are shown in Figure~\ref{fig:cache_bw}. We see that the maximum achieved bandwidth on a core is 73 GB/s, 96 GB/s, 52 GB/s, 69 GB/s for \texttt{scale1}, \texttt{scale2}, \texttt{saxpy1}, \texttt{saxpy2}, respectively. The bandwidth of \texttt{scale2} and \texttt{saxpy2} is 1.3$\times$ larger (than \texttt{scale1} and \texttt{saxpy1}, respectively) because the data cache allows a read/write cache-line replacement to happen in a single cycle~\footnote{\url{http://software.intel.com/en-us/articles/intel-xeon-phi-core-micro-architecture}}. The L1 bandwidth on a single core could be larger when further unrolling the loops or better scheduling instructions for each dataset. The aforementioned numbers are achieved by unrolling the loops 16 times without changing the assembly code. 

Furthermore, it is difficult to exactly measure the L2 bandwidth due to the presence of the L1 cache. The bandwidth depends on the memory access patterns. Specifically, when we use a L2-friendly memory access pattern, the compiler will identify the stream pattern and prefetch data to the L1 cache in time. By this, we will get a much larger bandwidth due to the common efforts of L1 and L2. On the other hand, an unfriendly memory access will experience many L1 misses and result in cache thrashing. Our benchmarking results are obtained when disabling the software prefetching. When using 4 threads on a core, we notice a bandwidth of 11 GB/s, 20 GB/s, 10 GB/s, 16 GB/s for \texttt{scale1}, \texttt{scale2}, \texttt{saxpy1}, \texttt{saxpy2}, respectively (Figure~\ref{fig:cache_bw}).

\begin{figure*}[!t]
\centering
\subfigure[Core Distribution]{\label{fig:core_mem_distribution}\includegraphics[width=0.35\textwidth]{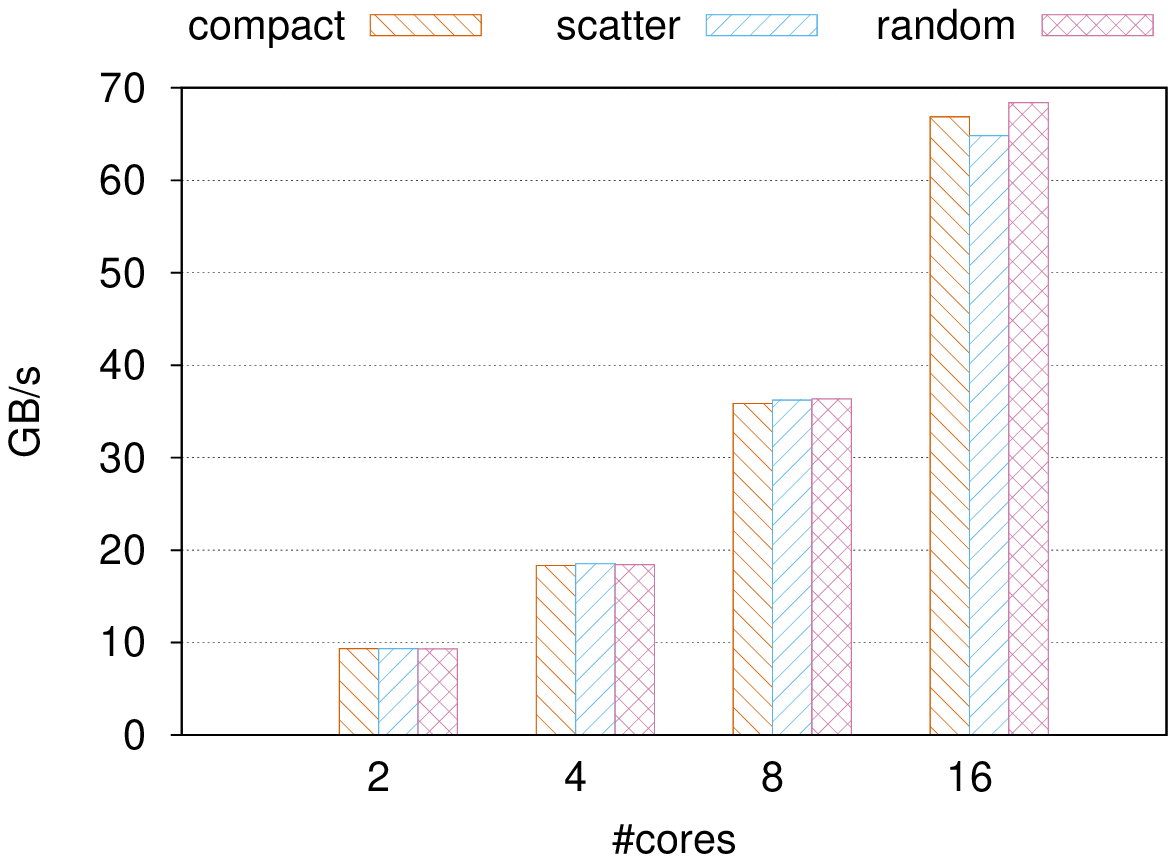}}
\subfigure[Thread Distribution]{\label{fig:thread_core_distribution}\includegraphics[width=0.35\textwidth]{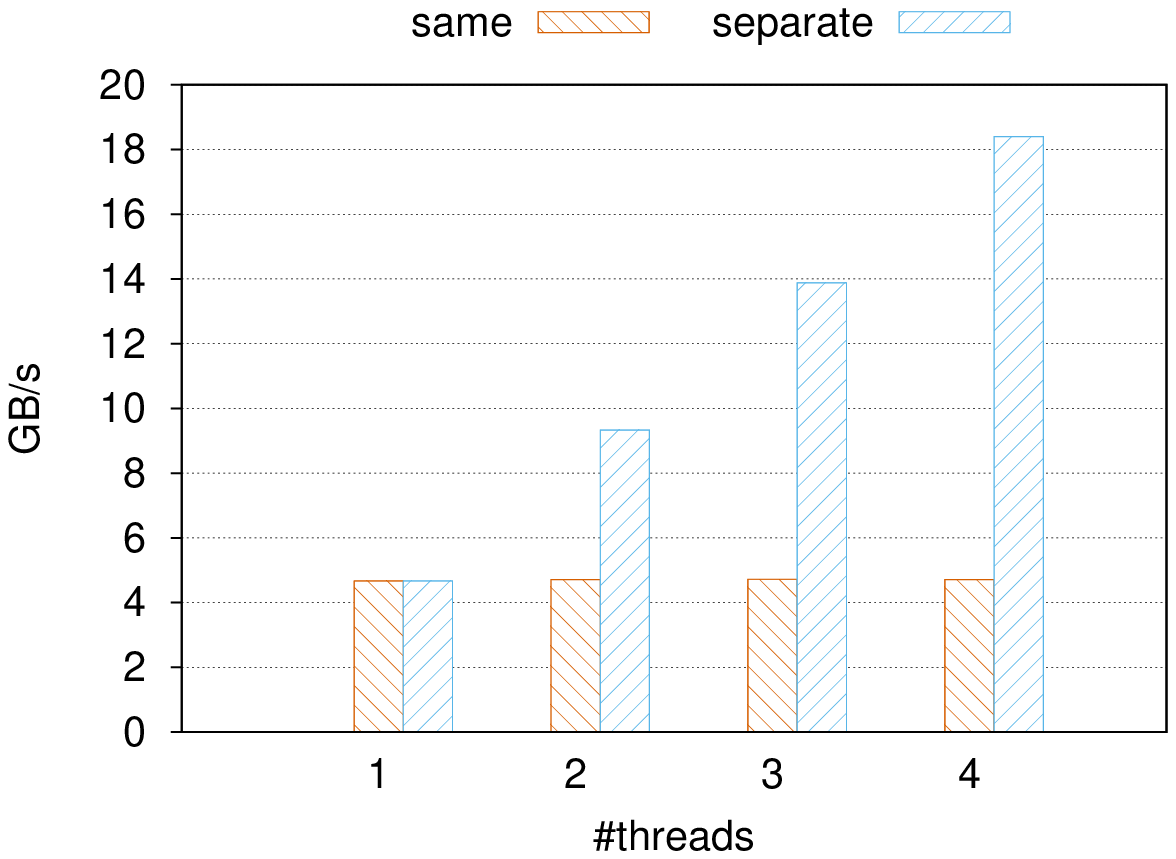}}
\caption{Core and thread distribution effects (we use the \texttt{read} kernel in Section~\ref{subsec:off_chip_mem_bw} and the array size is 1 GB).}
\label{fig:core_thread_distri_tot}
\end{figure*}

\subsection{Ring Interconnect}

On Xeon Phi, the cores and memory controllers are interconnected in a bi-directional ring. When multiple threads are requesting data simultaneously, shared components like the ring stop or DTDs can become performance bottlenecks. In order to check this hypothesis, and its eventual performance impact, we use \textit{thread affinity} to fix threads on cores, and we run the bandwidth microbenchmarks to quantify potential bandwidth changes (in GB/s) for different thread-to-core mapping scenarios. 

\subsubsection{Core/Thread Distribution} \label{subsec:ctde}
First, we measure the \texttt{read} memory bandwidth by distributing threads onto separate cores in three different patterns: (1) \textit{compact} - the cores are located close to each other, (2) \textit{scattered} - the cores are evenly distributed around the ring, and (3) \textit{random} - the core \texttt{IDs} are selected randomly with no repeats. 
The bandwidths are measured using 2, 4, 8, and 16 cores and the results are presented in Figure~\ref{fig:core_mem_distribution}. We see that the three approaches achieve very similar memory bandwidths. Thus, the cores around the ring are \textit{symmetric} on Xeon Phi, and the distance between has practically no impact on the achieved bandwidth. 

Second, as each Xeon Phi core supports up to four hardware threads, we investigate whether there is any impact on bandwidth if the threads are all gathered on the same core (thus, less interconnect traffic) or distributed among different cores. 
Figure~\ref{fig:thread_core_distribution} shows that when the threads run on the same core, the bandwidth stabilizes at 4.7 GB/s. 
We also note that running threads on separate cores results in a linear bandwidth increase with the number of threads. We conclude that when multiple threads on the same core request data simultaneously, they will compete for the shared hardware resources (e.g., the ring stops), thus serializing the requests.

\subsubsection{Accessing Shared-Data}
\begin{figure}[!t]
\centering
\includegraphics[width=0.35\textwidth]{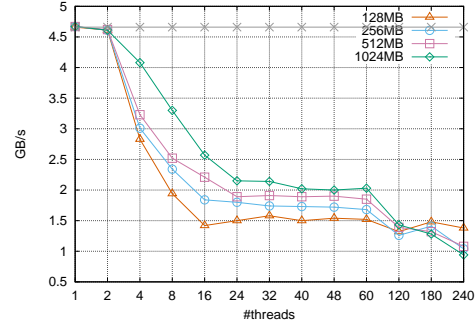}
\caption{The memory bandwidth when the threads read the same memory space. We use the \textit{compact} way to distribute threads onto cores.}
\label{fig:interco_same}
\end{figure}

Section~\ref{subsec:ctde} focuses on the achieved bandwidth when threads access separate memory spaces. In this section we investigate \textbf{what is the bandwidth when different threads access the same memory space simultaneously?} 
We expect that the bandwidth would resemble that obtained by a single thread, assuming the memory requests are served by \textit{broadcasting}. 
Figure~\ref{fig:interco_same} presents the measured bandwidth, showing that the \texttt{read} bandwidth decreases over the number of threads until 24 (or 16). Thereafter, the bandwidth is constant around 1.5-2.0 GB/s (i.e., one third of the single thread bandwidth). When using more threads than cores, the bandwidth drops even further. 
This behavior is different from the linear increase trend (shown in Figure~\ref{fig:core_mem_distribution}) seen when accessing separate memory spaces. We assume the bottleneck lies in the simultaneous access to the DTDs. Therefore, for bandwidth gain, applications should strive to keep threads accessing different parts/cachelines of the shared memory space (for as much as possible), to avoid the effects of contention at the interconnect level. 


\begin{figure*}[!t]
\centering
\subfigure[Host $ \Longrightarrow$ Phi]{\label{fig:offload_bw_snd}\includegraphics[width=0.35\textwidth]{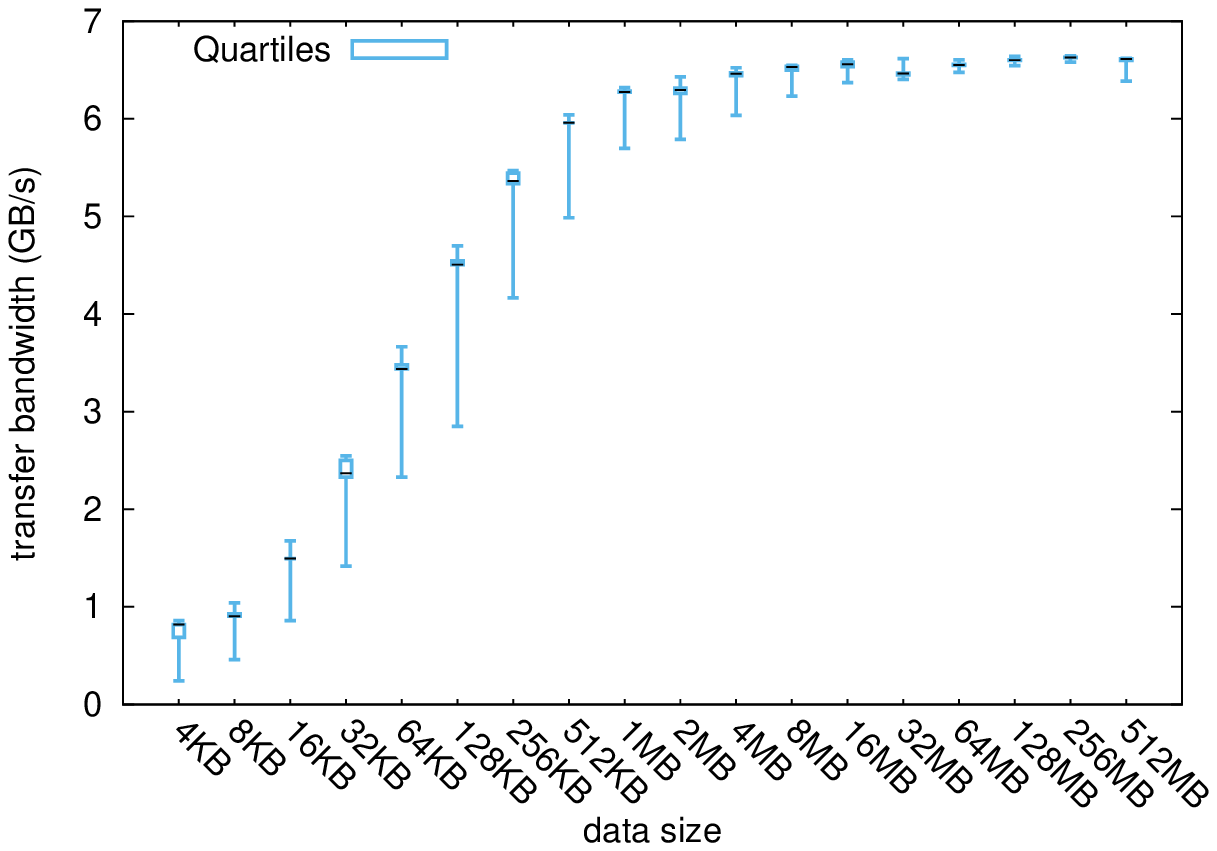}}
\subfigure[Phi $ \Longrightarrow$ Host]{\label{fig:offload_bw_rcv}\includegraphics[width=0.35\textwidth]{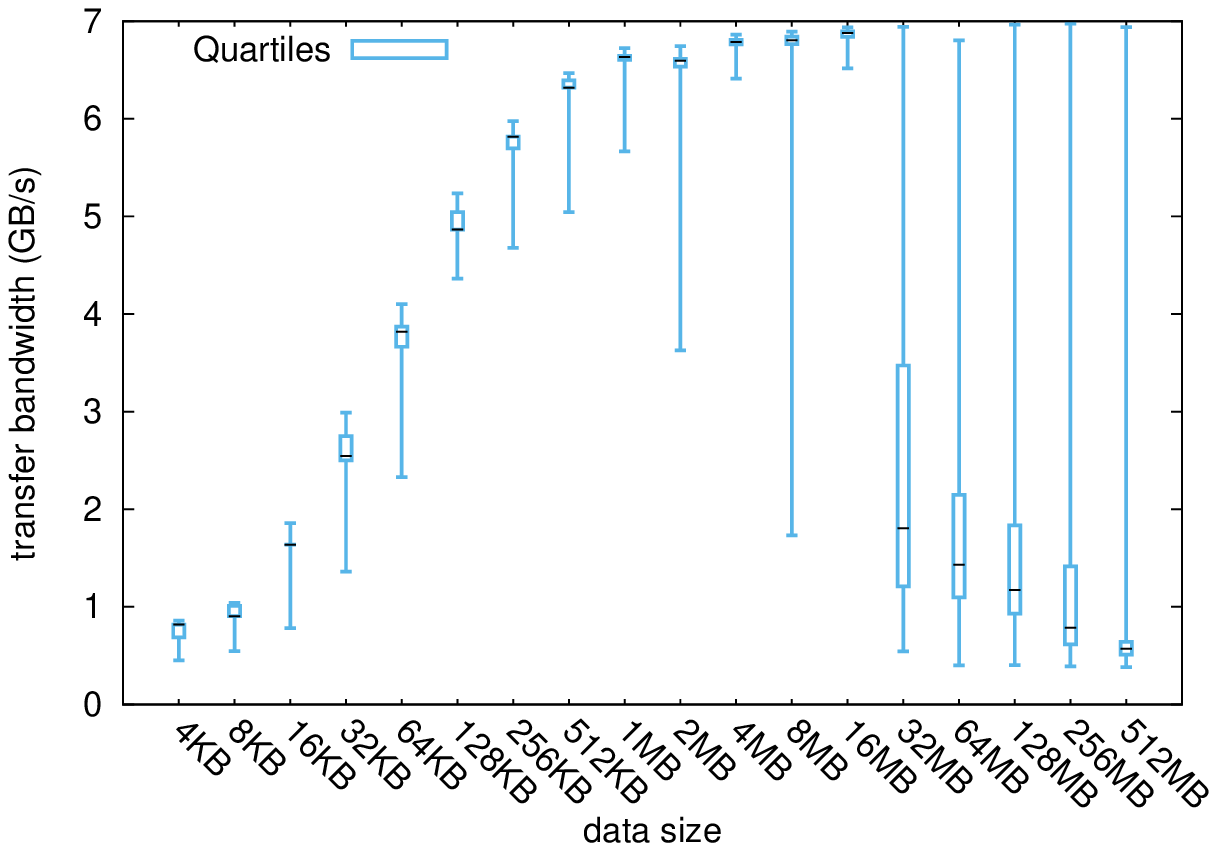}}
\caption{Achieved data transfer bandwidth (over PCIe) between a host and the Xeon Phi working as a coprocessor.}
\label{fig:offload_bw}
\end{figure*}

\subsection{PCIe Data Transfer}
When used as a coprocessor, Xeon Phi is connected via PCIe to a host (e.g., a traditional CPU). When offloading computation to the Xeon Phi, the tasks and the related data need to be transferred back and forth between the two processors. As seen for GPUs~\cite{citeulike:6569945}, these transfers can be expensive in terms of overall application performance. Thus, we have designed a benchmark to measure the data transfer bandwidth~\footnote{Our benchmark is based on Intel's sample code: \url{http://software.intel.com/en-us/articles/how-to-achieve-peak-transfer-rate}}. 

To do so, we use the \texttt{offload} pragma (specifying \texttt{in} and \texttt{out} for the transfer direction) to transfer datasets of different sizes from host to Xeon Phi and back. The transferred data is allocated with a 4K alignment, for optimal DMA performance~\cite{citeulike:12204757}. The achieved bandwidth between host and Xeon Phi is presented in Figure~\ref{fig:offload_bw} (we report the results over 1000 times). 
We note that the bandwidth increases with data size, and it is relatively stable for different runs, for both directions.
However, for data transfers larger than 32 MB, the Phi to host bandwidth shows a large variation, with the \texttt{median} bandwidth value decreasing sharply (up to 6 times!). 
The reasons for this large variance are still under investigation. 


\section{Summary and Comparison with SDG~\cite{citeulike:12204757}} \label{sec:comparison}

From our thorough benchmarking of Xeon Phi, we summarize the following observations: 

\begin{itemize}
\item \textbf{High Throughput: } Xeon Phi is indeed a high-throughput platform. The peak instruction throughput is achievable, but it depends on the following factors: (1) the number of threads and threads/core occupancy, (2) the utilization of the 512-bit vectors, (3) the issuing width (i.e., the number of independent instruction streams), (4) the instruction mix. Furthermore, single-precision data leads to better performance for math-intensive kernels. 

\item \textbf{Memory Selection:} Accessing the local L1 cache is 8 times faster than accessing the local L2 cache, which is again an order of magnitude faster than accessing the remote caches or the off-chip memory. However, the difference between a remote cache access and an off-chip memory access is relatively small (17\%). Furthermore, the remote access latency does not depend on the cache-line state. 
    
\item \textbf{Efficient Memory Access:} Data is read and written from/to the off-chip memory in cache lines (64 Bytes). The maximum achievable bandwidth is 164 GB/s for \texttt{read} operations and 76 GB/s for \texttt{write} operations - a lot lower than the theoretical peak of 320 GB/s. With \texttt{streaming store} instructions, the \texttt{write} bandwidth can increase up to 1.7 times. 
Further, programmers need many threads (at least 60 - one per core) to issue enough memory requests to saturate the ring interconnect and the memory channels. 
The hardware and software prefetching can improve bandwidth; their efficiency increases with the length of the \texttt{stanzas} of contiguous memory accesses.
Finally, disabling ECC leads to an average of 20\% increase in bandwidth. 

\item \textbf{Ring Interconnect:}  All cores can be seen as \textit{symmetrical} peers, and the distance between communicating cores has little impact on performance. However, memory requests from threads running on the same core are serialized, provided that the bandwidth reaches 4.7 GB/s. Furthermore, when threads (on different cores) are accessing the same data, the simultaneous access to the DTD  leads to bandwidth loss.
\end{itemize}

These observations are machine-centric optimizations. We also compare our results with the information provided by the Intel Software Development Guide (SDG) in Table~\ref{tbl:comparison}, and we note that we did improve on both the content and the accuracy of the official data: instruction latency data, local and non-local memory access bandwidth and latency data, an interconnect study, and a PCIe offload evaluation.  
We also show that the reported 1-TFlops performance is achievable while the 320-GB/s memory bandwidth seems not to be. 

Overall, we believe our results are complementary to the SDG, and, being backed up by more practical guidelines, be of added value for programmers using this platform. 

\begin{table}[htbp]
\caption{A comparison of our results with the data available in the SDG (`N/A' stands for "not available" in SDG).}
\begin{center}
\begin{tabular}{|c|c|c|}
\hline \hline
\textbf{Metric} & \textbf{SDG} & \textbf{Measured} \\ \hline
\multicolumn{3}{|l|}{VPU} \\ \hline
Latency & general statement & cycles/instruction \\ 
Throughput & 1008 GFlops & 1008 GFlops \\
EMU evaluation & general statement & quantified \\ \hline
\multicolumn{3}{|l|}{L1 Cache} \\ \hline
Latency (local) & 1 cycle & 3 cycles \\ 
bandwidth (local) & N/A & R=64B/c;W=64B/c \\ \hline
\multicolumn{3}{|l|}{L2 Cache} \\ \hline
Latency (local) & 11 cycles & 24 cycles \\ 
Bandwidth (local) & N/A & 12 GB/s \\ 
Latency (remote) & N/A & 250 cycles \\ \hline
\multicolumn{3}{|l|}{Off-chip memory} \\ \hline
Latency & N/A & 302 cycles \\ 
Bandwidth & 320 GB/s & R=164GB/s;W=76GB/s \\ 
Prefetching & general statement & quantified \\ 
ECC factor & general statement & quantified \\ \hline
\multicolumn{3}{|l|}{Interconnections} \\ \hline
Ring Traffic Contention & N/A & ring stops, DTDs \\ 
PCI Express Bandwidth & N/A & up to 7 GB/s \\ \hline \hline
\end{tabular}
\end{center}
\label{tbl:comparison}
\end{table}



\section{A Simplified View of Xeon Phi} \label{sec:model}

Based on our results and observations, we attempt to build a simplified view of the Xeon Phi, providing programmers with a simpler platform model for reasoning about parallel algorithm design. This platform model captures the key performance features of the processor, ensuring good performance with relatively low programming effort (i.e., using high-level programming tools). 

\begin{figure}[!t]
\centering
\includegraphics[width=0.40\textwidth]{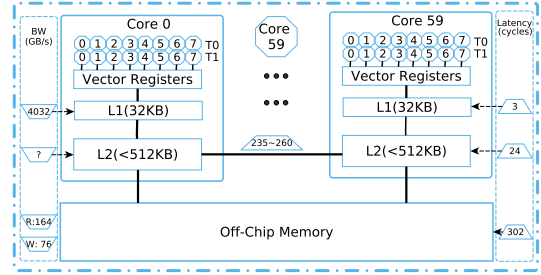}
\caption{A simplified view of Intel Xeon Phi.}
\label{fig:mic_view}
\end{figure}

Figure~\ref{fig:mic_view} shows the machine model for Xeon Phi. The machine has 60 symmetrical cores, each of which contains 2 vector threads working on 8 double-precision or 16 single-precision data elements in a lock-step manner. 
Each time a thread requests data elements from the main memory, it will load a cache-line (64 bytes) into its L1 and L2 cache - thus, memory bandwidth is improved if all the processing elements of a vector thread access the data elements located in the same cache-line. Furthermore, compared with accessing local caches, remote caches and off-chip memory are slow (see the numbers in the figure)

This machine model limits itself to those architectural details that are important for performance. For example, programmers do not have to keep the ring interconnect in mind because the cores perform symmetrically. On the other hand, this model plus the optimization guidelines provide enough detailed information for modeling and optimizing applications. Our initial results show the power of the machine model in guiding kernel design~\cite{myself_bm_phi}, and we are now working on a quantified performance model of Xeon Phi based on it. 


\section{Related Work} \label{sec:bm_relwk}
In this section, we survey and briefly discuss the work related to our (micro)benchmarking approach. We focus mainly on existing CPU and GPU benchmarking methods, as there are no other comprehensive studies of Xeon Phi - yet.  

In~\cite{citeulike:9482128}, the authors develop a high-level program to evaluate the cache and TLB for any machine. Part of our work 
is based on their approaches (targeting uni-core processors, though).
Multiple studies are also performed on multi-core CPUs. In~\cite{citeulike:12186871}, the authors report performance numbers from three multi-core processors 
, including not only execution time and throughput, but also a detailed analysis on the memory hierarchy performance and on the performance scalability between single and dual cores. 
Daniel Molka et al.~\cite{citeulike:8778343} revealed many fundamental details of the Intel Nehalem using benchmarks for latency and bandwidth between different locations in the memory subsystem. We use similar approaches for the access latency of remote caches.

For GPUs, Volkov et al.~\cite{citeulike:6569945} presented detailed benchmarking of the GPU memory system that reveals sizes and latencies of caches and TLB. 
Later, Wong et al.~\cite{citeulike:9219351} presented an analysis of the NVIDIA GT200 GPU and their measurement techniques. 
They used a set of micro-benchmarks to reveal architectural details of the processing cores and the memory hierarchies. Their results revealed the presence of some undocumented hardware structures. 
While these microbenchmarks are in CUDA and targeted NVIDIA GPUs, Thoman et al.~\cite{citeulike:12298292} develop a set of OpenCL benchmarks targeting a large variety of platforms. They include code designed to determine parameters unique to OpenCL, like the dynamic branching penalties prevalent on GPUs. They also demonstrate how their results can be used to guide algorithm design and optimization

Garea et al.~\cite{id_ramos_hpdc} developed an intuitive performance model for cache-coherent architectures and demonstrated its use on Intel Xeon Phi. Their model is based on latency measurements, which match well with our latency results. In addition to the cache access latency, we have shown how we benchmark the instruction throughput, the memory bandwidth at different levels, and the interconnect performance.

\section{Conclusion and Future Work} \label{sec:concld}
Given its performance promises, Intel Xeon Phi is very likely to become popular in the next generation of supercomputers. Therefore, our work focused on providing a benchmarking strategy and several key insights into the performance of this new many-core processor. 
By using a set of microbenchmarks, we characterized the major components of this architecture - cores, memory, and interconnections - summarized into four machine-centric observations (and potential optimization guidelines). We also made a first attempt to provide a simplified machine view to facilitate application design and performance tuning on Xeon Phi. 

In general, our benchmarking results are consist with Xeon Phi's published data. However, the data we have added through this benchmarking effort allowed us to expose more accurately the expected key performance factors for the Xeon Phi. We have shown that the platform is able to deliver its performance promises in terms of computation, but programmers will need to find the right parallelization strategy to fill 240 hardware threads with compute-intensive tasks, while finding the right balance between data partitioning and coherent memory requests to achieve sufficient memory bandwidth. Thus, we believe the number of applications that can easily use Xeon Phi's potential in their existing, naive form is, for now, very limited. And for high performance, programmers need to take a lot of efforts on parallelization, analysis, and optimization~\cite{citeulike:12608371, citeulike:12608375}.

In terms of future work, we are working on a quantified performance model for Xeon Phi, which could be used in identifying performance bottlenecks and guiding performance optimization. This potential model should start with the microbenchmarks and application characteristics.

\section*{Acknowledgment}
The authors would like to thank Evghenii Gaburov from SURFsara for the on-line discussions on cache bandwidth. This work is partially funded by CSC (China Scholarship Council), and the National Natural Science Foundation of China under Grant No.61103014 and No.11272352.

\bibliography{mybib}{}
\bibliographystyle{ieeetr}

\end{document}